\documentclass{aa}
\usepackage{graphicx}
\usepackage{txfonts}
\def\ltsima{$\; \buildrel < \over \sim \;$}
\def\lsim{\lower.5ex\hbox{\ltsima}}
\def\gtsima{$\; \buildrel > \over \sim \;$}
\def\gsim{\lower.5ex\hbox{\gtsima}}

\newcommand{\be}{\begin{equation}}
\newcommand{\en}{\end{equation}}

\def\cmdue {\rm \ cm^{-2}}

\begin{document} 
\title{Evidence for intrinsic absorption in the Swift X--ray afterglows}

\author{
S.~Campana\inst{1}
\and
P. Romano\inst{1}
\and
S. Covino\inst{1}
\and
D. Lazzati\inst{2}
\and 
A. De Luca\inst{3}
\and
G. Chincarini\inst{1,4}
\and
A. Moretti\inst{1}
\and 
G. Tagliaferri\inst{1}
\and
G. Cusumano\inst{5}
\and
P. Giommi\inst{6}
\and
V. Mangano\inst{5}
\and
M. Perri\inst{6}
\and
V. La Parola\inst{5}
\and
M. Capalbi\inst{6}
\and
T. Mineo\inst{5}
\and
L.A. Antonelli\inst{7}
\and
D.N. Burrows\inst{8}
\and
J.E. Hill\inst{8}
\and
J.L. Racusin\inst{8}
\and
J.A. Kennea\inst{8}
\and
D.C. Morris\inst{8}
\and
C. Pagani\inst{8,1}
\and
J.A. Nousek\inst{8}
\and
J.P. Osborne\inst{9}
\and
M.R. Goad\inst{9}
\and
K.L. Page\inst{9}
\and
A.P. Beardmore\inst{9}
\and
O. Godet\inst{9}
\and
P.T. O'Brien\inst{9}
\and
A.A. Wells\inst{10}
\and 
L. Angelini\inst{11,12}
\and
N. Gehrels\inst{11}
}

\offprints{S. Campana, campana@merate.mi.astro.it}

\institute{INAF - Osservatorio Astronomico di Brera, Via Bianchi 46, I-23807
Merate (LC), Italy  
\and
JILA, Campus Box 440, University of Colorado, Boulder, CO 80309-0440, USA
\and
INAF - Istituto di Astrofisica spaziale e Fisica Cosmica, Via Bassini
15, I-20133, Milano, Italy   
\and
Universit\'a di Milano-Bicocca, Dipartimento di Fisica, Piazza della Scienza
3, I-20126 Milano, Italy
\and
INAF - Istituto di Astrofisica spaziale e Fisica Cosmica, Via La Malfa 153,
I-90146 Palermo, Italy
\and
Agenzia Spaziale Italiana, Science Data Center, Via Galileo Galilei,
I-00044, Frascati, Italy
\and
INAF - Osservatorio Astronomico di Roma, Via di Frascati 33, I-00040
Monteporzio Catone (Roma), Italy
\and
Department of Astronomy and Astrophysics, 525 Davey Lab., Pennsylvania State
University, University Park PA 16802, USA
\and
Department of Physics and Astronomy, University of Leicester, Leicester LE1
7RH, UK
\and
Space Research Centre, University of Leicester, Leicester LE1
7RH, UK
\and 
NASA/Goddard Space Flight Center, Greenbelt MD 20771, USA
\and 
Department of Physics and Astronomy, Johns Hopkins University, 3400 North
Charles Street, Baltimore MD 21218, USA
}

\date{Received; accepted}

\abstract{Gamma-ray burst (GRB) progenitors are observationally linked to
the death of massive stars. X--ray studies of the GRB afterglows can deepen
our knowledge of the ionization status and metal abundances of the matter in
the GRB environment. Moreover, the presence of local matter can be inferred
through its fingerprints in the X--ray spectrum, i.e. the presence of
absorption higher than the Galactic value. A few studies based on BeppoSAX
and XMM-Newton found evidence of higher than Galactic values for the column
density in a number of GRB afterglows. Here we report on a systematic analysis
of 17 GRBs observed by Swift up to April 15, 2005. We observed a large number
of GRBs with an excess of column density. Our sample, together with previous
determinations of the intrinsic column densities for GRBs with known redshift,
provides evidence for a distribution of absorption consistent with that
predicted for randomly occurring GRB within molecular clouds.
}

\titlerunning{Intrinsic $N_H$ in Swift GRBs}
\authorrunning{S. Campana et al.}

\maketitle

\section{Introduction}
\label{intro}

Evidence has been accumulating in recent years that at least a subclass of
Gamma--ray bursts (GRBs), the ones with a long ($\gsim 2$ s) burst event, are
associated with deaths of massive stars (e.g. Woosley 1993; Paczy\'nski 1998;
MacFadyen \& Woosley 1999). This evidence was initially based on the
relatively small offset of the GRB location with respect to the center of the
host galaxy (Bloom, Kulkarni \& Djorgovski 2002). Moreover, decisive
supernova features have been observed in the afterglow of a few nearby GRBs
(Galama et al. 1998, Della Valle et al. 2003; Stanek et al. 2003;
Malesani et al. 2004), directly linking long GRBs to massive stars. This also
provides strong observational evidence for the connection of GRBs to star
formation (Djorgovski et al. 1998; Fruchter et al. 1999; Prochaska et
al. 2004). A study on the GRB host galaxies 
by Le Floc'h et al. (2003) found that these hosts have very blue colours,
comparable to those of the faint blue star-forming sources at high
redshift. The association of long GRBs with star forming regions supports
the idea that a large fraction of the optically-dark GRBs (i.e. GRBs without
an optical afterglow), as well, are due to high (dust) absorption
(Lazzati, Covino \& Ghisellini 2002; Rol et al. 2005; Filliatre et al. 2005).  

Together with optical studies, which probe the dust content of the GRB
environment, X--ray studies of the GRB afterglows can give insight on the  
ionization status and metal abundances of the matter in the GRB environment. 
This can be done by using either emission or absorption features (e.g.,
B\"ottcher et al. 1999; Ghisellini et al. 2002). Although emission features are
more apparent, the cumulative effect of low-energy cutoff is easier to
detect in the relatively low signal to noise spectra of X--ray afterglows. 
Moreover, if the absorbing material is located close to the GRB site ($\sim
0.1-10$ pc), it is expected that GRB photons may lead to a progressive
photoionization of the gas, gradually reducing the effect of low energy
absorption (Lazzati \& Perna 2002; Perna \& Lazzati 2002). 

X--ray absorption in excess of the Galactic value has been
reported for a handful of GRB afterglows (Owens et al. 1998; Galama \& Wijers
2001; Stratta et al. 2004; De Luca et al. 2005; Gendre, Corsi \& Piro 2005). 
Evidence for a decrease of the intrinsic column density with time in the
X--ray prompt emission of some GRBs has also been found (GRB980506, Connors \&
Hueter 1998; GRB980329, Frontera et al. 2000; GRB011211, Frontera et
al. 2004). Lazzati \& Perna (2002) interpreted them as evidence for GRBs
occurring within overdense regions in molecular clouds similar to star
formation globules in our Galaxy. 

Stratta et al. (2004) presented a systematic analysis of a sample of 13 bright
afterglows observed with BeppoSAX narrow field instruments. They found a
significant detection of additional intervening material in only two cases
(namely, GRB990123 and GRB010222), but, owing to the limited photon
statistics, they could not exclude that intrinsic X--ray absorption is also
present in the other bursts.  
Chandra observations of GRB afterglows have yielded a few detections and
constraints of the presence of intrinsic X--ray absorption (Gendre et
al. 2005). XMM-Newton observed 9 GRB
afterglows (for a review see De Luca et al. 2005 and Gendre et
al. 2005). These are mainly INTEGRAL GRBs, the large majority of which has
been discovered close to the Galactic plane (i.e. are characterized by a
relatively high Galactic column density). From XMM-Newton data one can
gather evidence that at least several GRBs occur in high density regions
within their host galaxies (e.g, De Luca et al. 2005).

Here we investigate the presence of intrinsic absorption in the complete set
of all the 17 GRBs promptly observed by Swift up to April 15, 2005.
The paper is organized as follows. In section 2 we present the data collected
by Swift and the analysis procedure. In section 3, we derive values and/or
upper limits on the column density in excess of the Galactic value (based on
the maps by Dickey \& Lockman 1990). For 6 GRBs a redshift has also been
determined through spectroscopic observations. For these we investigate the
intrinsic excess of absorption. In section 4 we discuss our results.
Section 5 is dedicated to our conclusions.  

\section{Swift data}
\label{data}

The X--Ray Telescope (XRT, Burrows et al. 2005a) on board Swift (Gehrels et
al. 2004) is a focusing X--ray telescope with a 110 cm$^{2}$ effective
area at 1.5 keV, 23 arcmin field of view, 18 arcsec resolution (half-power diameter) and 
0.2--10 keV energy range. The first GRB followed by XRT was GRB041223 (Burrows
et al. 2005b). Since then 22 other GRBs were observed by Swift up to April
15, 2005, together with one discovered by HETE II.
Of these 24 GRBs, 19 were observed by XRT and only two of them [GRB050410 and
GRB050117a (see also Hill et al. 2005)] were not detected or have
too few photons to perform a meaningful spectral analysis.
For eight of them Swift was able to repoint within a
few hundred seconds, whereas the remaining nine were observed at later
times ($>30$ min). In Table 1 we present a log of the XRT observations  
used in the present work. 

 \begin{table}
\label{datalog}
\caption{Observation log.}
\begin{tabular}{ccccc}
\hline
GRB    & Start time&Mode& Region$^{\$}$& Exp. time\\
       & (s)$^*$   &    &        & (s)\\
\hline
041223 & 16661	   & PC & 0, 20 & 4018 \\
050124 & 11113     & WT & 40x20 & 7351 \\
       &           & PC & 0, 20 & 11066\\
050126 &   131     & PC & 2, 20 & 278 \\
       &           & PC & 0, 13 & 8720 \\
050128 &   108     & PC & 5, 20 & 2878 \\
       &           & PC & 0, 15 & 14316 \\
050215b&  2100     & PC & 0, 10 & 35563 \\
050219a&    92     & WT & 40x20 & 5003 \\
       &           & PC & 0, 20 & 2570  \\
050219b&  3130     & WT & 40x20 & 15051 \\
       &           & PC & 0, 20 & 5724 \\
050223 &  2875     & PC & 0, 10 & 2337 \\
050306 &127390     & PC & 0, 10 & 12284 \\
050315 &    83     & WT & 40x20 & 748  \\
       &           & PC & 7, 40 & 690 \\
       &           & PC & 0, 20 & 9642 \\
050318 &  3277     & PC & 2, 20 & 395 \\
       &           & PC & 0, 20 & 3938 \\
050319 &    87     & PC & 3, 40 & 1495 \\
       &           & PC & 0, 40 & 2511 \\
050326 &  3258     & PC & 1, 16 & 60 \\
       &           & PC & 0, 15 & 40723\\ 
050401 &   131     & WT & 40x20 & 2292 \\
       &           & PC & 3, 20 & 276\\
       &           & PC & 0, 15 & 2079\\
050406 &    92     & WT & 40x20 & 109 \\
       &           & PC & 2, 30 & 70 \\
       &           & PC & 0, 10 & 49050 \\  
050408 &  4653     & PC & 0, 11 & 1041 \\
       &           & PC & 0, 5  & 57532 \\
050412 &    89     & WT & 40x20 & 168 \\ 
       &           & PC & 2, 30 & 145 \\
\hline
\end{tabular}

$^*$ Time from the BAT trigger time.

$^{\$}$ Extraction region. In PC mode it is a circular/annular region with inner
and outer radii reported in the column. In the case of WT mode the region is
always a $40\times20$ region along the column centered on source. 
\end{table}

\begin{table}
\label{values}
\caption{Results of the spectral analysis.}
\begin{tabular}{cccc}
\hline
GRB    &$N_H$ Gal.$^a$        & $N_H$ obs.$^b$   & $\chi^2_{\rm red}$ \\
       &$10^{20}\cmdue$         & $10^{20}\cmdue$    & (dof)	      \\
\hline
041223 & 10.9 (9.9) [5.6]    &$16.8^{+5.2}_{-4.2}$ & 0.8 (26)      \\
050124 & 5.2 (2.6) [1.7]     &$6.2^{+3.9}_{-2.5}$  & 1.3 (35)      \\
050126 & 5.3 (3.2) [2.6]     &$4.1^{+2.7}_{-2.6}$  & 0.9 (10)      \\
050128 & 4.8 (4.9) [3.8]     &$12.5^{+1.4}_{-1.3}$ & 1.3 (105)     \\
050215b& 2.1 (2.0) [0.9]     &$<3.4$            & 1.0 (4)       \\
050219a& 8.5 (10.1) [8.1]    &$30.1^{+6.5}_{-5.9}$ & 1.0 (57) \\     
050219b& 3.8 (3.0) [1.7]     &$23.8^{+4.0}_{-3.7}$ & 1.0 (98)     \\
050223 & 7.1 (6.6) [4.4]     &$9.5^{+23.8}_{-7.3}$ & 1.2 (3)      \\
050306 & 3.1 (2.9) [3.5]     &$46.1^{+35.6}_{-28.8}$& 1.0 (3)      \\
050315 & 4.3 (3.3) [2.5]     &$14.9^{+3.9}_{-2.2}$ &  1.3 (42)     \\
050318 & 2.8 (1.8) [0.9]     &$4.2^{+1.9}_{-1.5}$  & 0.8 (39)      \\
050319 & 1.1 (1.2) [0.5]     &$3.0^{+0.9}_{-0.8}$  & 1.3 (29)      \\
050326 & 4.5 (3.8) [1.8]     &$18.9^{+7.1}_{-6.0}$ & 1.0 (25)      \\
050401 & 4.8 (4.4) [3.3]     &$21.1^{+2.2}_{-1.8}$ & 1.0 (297)     \\
050406 & 2.8 (1.7) [1.1]     &$<6.6$            & 1.2 (10)      \\
050408 & 1.7 (1.5) [1.3]     &$30.7^{+5.5}_{-4.9}$ & 0.9 (45)   \\
050412 & 2.2 (1.7) [1.0]     &$26.4^{+14.9}_{-12.4}$& 1.4 (11)    \\ 
\hline
\end{tabular}

$^a$ Column density values are from Dickey \& Lockman (1990). Values between
parentheses are from Kalberla et al. (2005) and in square parentheses from
Schlegel, Finkbeiner \& Davis (1998), using the usual conversion
$N_H=5.9\times 10^{21}\,E(B-V)\cmdue$.  

$^b$ The values of the column density have been computed at $z=0$ since we do
not have any knowledge of the GRB redshift for most of them (but see Table \ref{zz}).
\end{table}

GRBs are observed by XRT with different observing modes and source count
rates. These modes were designed to minimize photon pile-up when observing
the highly variable flux of GRB afterglows. The change between observing modes
should occur automatically when the XRT is in the so-called Auto State (for a
thorough description of XRT observing modes see Hill et al. 2004).
Many of these early bursts were observed in Manual State instead, with the
observing mode fixed, during the calibration phase (before April 5, 2005). For
these GRBs observations were often carried out in Photon Counting (PC, the usual
mode providing 2D images) and for some bright bursts the initial data are
piled-up.  
This effect can be corrected by extracting light curves and spectra from an
annular region around the source center (rather than a simple circular region),
with a hole size that depends on the source brightness. As the afterglows
decays the pile-up effect becomes negligible and extraction from a circular
region is feasible. For bursts observed in Auto State observations started in
Window Timing (WT) mode (providing just 1D imaging). 
Cross-calibration between modes assures that the two modes, PC and WT, provide
the same rate (within a few percent) on steady sources. 

Here we analyzed the dataset shown in Table 1. All data were processed with
the standard XRT pipeline within 
FTOOLS 6.0 ({\tt xrtpipeline} v. 0.8.8) in order to produce screened event
files. WT data were extracted in the 0.5--10 keV energy range, PC data in the
0.2--10 keV range. Standard grade filtering was adopted (0--2 for WT and
0--12 for PC, according to XRT nomenclature, see Burrows et al. 2005a). From these we
extracted spectra using regions selected to avoid pile up and to maximize the
signal to noise (see Table 1). In WT mode we adopted the standard
extraction region of $40\times 20$ pixels along the WT line. In PC we used an 
annular region when the inner core of the Point Spread Function (PSF) was
piled-up and circular regions otherwise. The size of the extraction region
depends on the source strength and background level. Appropriate ancillary response
files were generated with the task {\tt xrtmkarf}, accounting for PSF
corrections. The latest response matrices (v.007) were used. The data were
rebinned to have 20 counts per energy bin (in some cases with few photons
energy bins with as low as 10 counts per bin were used). 

\begin{figure}
\centering
\includegraphics[height=8cm,angle=-90]{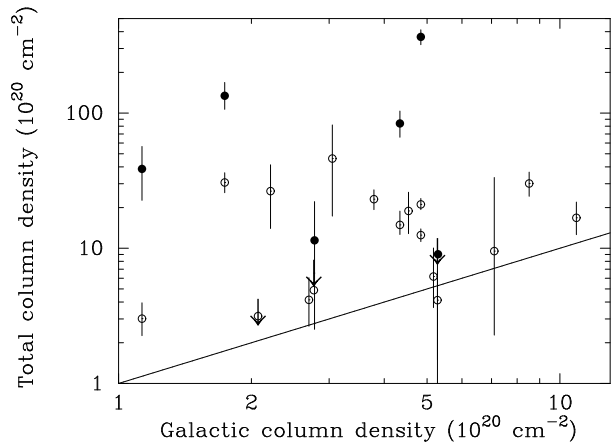}
\caption{Galactic column density versus column density obtained from spectral
fit of the X--ray afterglow. Open circles are values obtained without any
redshift information. Filled circles indicate values for the six GRBs with
known redshift at the redshift of the host galaxy. Upper limits are also
indicated with filled and open circles, as above. The line represents the
prints of equal values between the Galactic and the total column density
(i.e. no intrinsic absorption).
}
\label{nh}
\end{figure}

The data were fitted with a simple absorbed power law model. We have adopted
the usual photoelectric absorption using Wisconsin (Morrison \& McCammon 1983)
cross-sections ({\tt wabs} model within XSPEC). Normalizations were
left free to vary. Power law photon index and absorbing column densities were
first tied across the different observations and observing modes, but if the
fit was not satisfactory the photon index was allowed to vary between
observations (indicating a spectral evolution). We searched for variations of
the column density with time by allowing the column density to vary across
different observations.
For GRBs with known redshift we also fitted an absorbed power law model with a
fixed Galactic column density component plus a free column density at the redshift
of the GRB. Results are shown in Tables 2 and 3 and in
Fig. \ref{nh}, where GRB total column densities are plotted against the Galactic
column densities. 

\begin{table*}
\label{zz}
\caption{Results of the spectral analysis of GRB with known redshift.}
\begin{center}
\begin{tabular}{cccccc}
\hline
GRB    &Redshift & $N_H$ Gal.  & $N_H$ obs.$^*$   & $\chi^2_{\rm red}$ & $N_H$
DLA ($90\%$)\\
       & (ref.)  &$10^{20}\cmdue$& $10^{20}\cmdue$    & (dof)	          & $10^{20}\cmdue$  \\
\hline
050126 &1.29 (1) & 5.3       &$<9.4          $    & 1.0 (10)         & 0.8 (6.4)\\
050315 &1.95 (2) & 4.3	     &$83.7^{+19.7}_{-17.4}$ &  1.4 (43)      & 1.4 (5.7)\\
050318 &1.44 (3) & 2.8       &$11.3^{+10.7}_{-8.9}$ & 0.6 (39)        & 0.9 (12.9)\\
050319 &3.24 (4) & 1.1       &$38.6^{+18.0}_{-16.0}$& 1.4 (29)        & 3.6 (447)\\
050401 &2.90 (5) & 4.8       &$366^{+47}_{-46}$    & 1.1 (294)      & 3.0 (272)\\
050408 &1.24 (6) & 1.7       &$134^{+34}_{-28}$    & 1.0 (45)       & 0.8 (4.7)\\
\hline
980703 &0.97     & 5.8       &$290^{+71}_{-27}$    &            & 0.6 (1.7) \\ 
990123 &1.60     & 2.1       &$30^{+70}_{-20}$   	&            & 1.1 (23.1) \\ 
990510 &1.63     & 9.4       &$160^{+19}_{-13}$    &            & 1.1(25.1)\\ 
000210 &0.85     & 2.5       &$50^{+10}_{-10}$    &             & 0.0 (0.0) \\
000214 &0.47     &5.8        &$<2.7$            &            & 0.0 (0.0) \\
000926 &2.07     & 2.7       &$40^{+35}_{-25 }$    &            & 1.6 (73.9)\\
010222 &1.48     & 1.6       &$120^{+70}_{-60 }$   &             & 1.0 (15.1)\\
001025a&1.48     & 6.1       &$66^{+30}_{-30}$    &             & 1.0 (15.1) \\
020322 &1.80     & 4.6       &$130^{+20}_{-20}$    &            & 1.3 (40.7)\\
020405 &0.70     & 4.3       &$47^{+37}_{-37}$     &            & 0.0 (0.0) \\
020813 &1.25     &7.5        &$<36.5$           &            & 0.8 (5.0)\\
021004 &2.33     &4.3        &$<34$             &            & 2.0 (118)\\
030226 &1.98     & 1.6       &$68^{+41}_{-33}$     &            & 1.5 (60.8)\\
030227 &3.90     & 22        &$680^{+18}_{-38}$    &            & 5.2 (649)\\
030328 &1.52     &4.3        &$<44.3$           &            & 1.0 (17.3)\\
\hline
\end{tabular}
\end{center}

$^*$ Column density values have been computed at the GRB redshift.

Refs.: 1) Berger, Cenko, Kulkarni 2005; 2) Kelson \&  Berger 2005; 3) Berger
\& Mulchaey 2005; 4) Fynbo et al. 2005; 5) Fynbo et al. 2005; 6) Berger,
Gladders \& Oemler 2005. 
Values in the second part of the table are from Stratta et al. (2004), De Luca
et al. (2005) and Gendre et al. (2005).
\end{table*}


\section{Discussion}
\label{discu}

We have analyzed the X--ray spectra of 17 GRB afterglows observed with Swift
up to April 15, 2005. In at least 10 of them we find significant evidence that
the observed column density is higher than the Galactic value. In 
contrast to previous investigations (De Luca et al. 2005; Stratta et
al. 2004; Gendre et al. 2005) based on BeppoSAX, INTEGRAL and HETE II, the
Swift sample has GRBs at low Galactic extinction (all below $\lsim
10^{21}\cmdue$), therefore more effectively probing the presence of absorption
due to the GRB environment or host galaxy. 

The evidence that a large fraction of GRBs is characterized by an absorbing
column density larger than the Galactic one clearly points to a high density
interstellar medium in the proximity of the GRB (in fact with X--rays we
directly probe the GRB line of sight, whereas in the optical the observations
might be contaminated by the host galaxy contribution). 
Dense environments in the host galaxy, possibly associated with star forming
regions, provide a further clear signature in favour of the association
of long GRBs to the death of massive stars.
For GRBs characterized by a low number of counts no firm conclusions can be
drawn. Moreover, the effect of an intrinsic column density can be hidden
either by a large Galactic absorber (as often occurs for INTEGRAL GRBs) or by
a large redshift shifting the energy scale by $(1+z)$ and the column density
effective value by $\sim (1+z)^{2.6}$. 

\begin{figure}
\centering
\includegraphics[height=8cm,angle=-90]{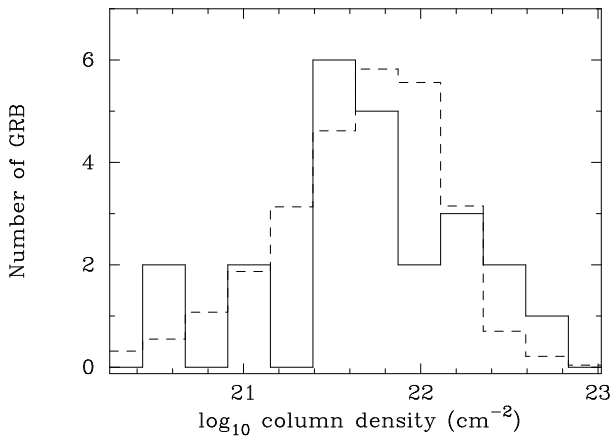}
\caption{Distribution (solid histogram) of intrinsic column density in a sample of
18 GRBs observed by Swift and other satellites (Stratta et al. 2004; De Luca
et al. 2005). This is compared with the expected distribution of column
density for GRB that occur in Galactic-like molecular clouds (dashed
histogram, from Reichart \& Price 2002).
}
\label{histo}
\end{figure}

We combine our sample of GRBs with known redshift with other intrinsic column densities
available in the literature (Stratta et al. 2004; De Luca et al. 2005; Gendre
et al. 2005), obtaining a sample of 21 GRBs (see Table \ref{zz}). In principle
the absorption excess found in most GRBs might not be local to the host galaxy
but may come from a line-of-sight interlooper. This often occurs in optical
studies of quasar with dumped Lyman absorber (DLA). Based on quasar studies
(Wolfe, Gawiser \& Prochaska 2005; P\'eroux et al. 2003) we simulated for each
of our bursts a distribution of line-of-sights (10000 trials), 
evaluating a mean absorption (weighted as $(1+z)^{2.6}$) and $90\%$ confidence
value. These values are reported in column six of Table \ref{zz}. The
influence of DLA systems is marginal in our sample even if there are a few GRBs
in which the observed absorption excess may come from intervening DLAs.
Indeed, such systems have recently been found in few Swift GRBs
(e.g. GRB050730 with $\log(N_H)=22.3$, Starling et al. 2005, Chen et al. 2005,
and GRB 050401, Watson et al. 2005, $\log(N_H)=22.5$). However, they are due
to the interstellar medium in the GRB host. One of these GRBs is part of our
sample, GRB 050401, and indeed we obtained a high value of the instrinsic column
density. 

In order to understand the origin of the absorption excess, 
we compared the distribution of measured intrinsic column densities with the
distribution expected for bursts occurring in Galactic-like molecular clouds
(Reichart \& Price 2002) and with the one expected for bursts
occurring following a host galaxy mass distribution using the Milky Way as a
model (Vergani et al. 2004). For each of 
these two column density distributions we simulated 10000 GRBs and compared,
by means of a Kolmogorov-Smirnov (KS) test, their instrinsic absorption
distribution to the observed distribution. 
We found that the observed distribution is inconsistent with
the galaxy column density distribution (Vergani et al. 2004), with a KS null
hypothesis probability of $10^{-11}$, but it is consistent with GRBs
distributed randomly in molecular clouds (KS null hypothesis probability of 0.61).
The simulated distribution in the case of bursts occurring in Galactic-like
molecular clouds and the observed one are plotted in Fig. \ref{histo}. This results
would support an origin of long GRBs within high density regions such as
molecular clouds. 
We stress that our sample also contains upper limits and that we are
sensitive to low values of column density, which however are found only in a
small fraction of the total sample.

\section{Summary and conclusions}

The main goal of the present paper is to investigate the presence of
intrinsic absorption in the X--ray spectra of GRB afterglows.
We analyzed a complete set of 17 afterglows observed by Swift XRT before
April 15, 2005. In 10 of them we found clear
signs of intrinsic absorption, i.e. with a column density higher than the
Galactic value estimated from the maps by Dickey \& Lockman (1990) at $>90\%$
confidence level (and with low probability of contamination from intervening
DLA systems).  
For the remaining 7 cases, the statistics are not good enough to draw firm
conclusions. This clearly suggests that long GRBs are associated with high
density regions of the interstellar medium, supporting the idea
that they are related to the deaths of massive stars. 

For the 6 GRBs with known redshift, together with 15 already known, we
can have an unbiased view of the intrinsic absorption in the host galaxy rest
frame. We found a range of $(1-35)\times 10^{21}\cmdue$ for all
GRBs. This range of values is consistent with the hypothesis that GRBs occur
within giant molecular clouds, spanning a range of column density depending on
their exact location (Reichart \& Price 2002). In our rest frame this column
density is then reduced by a factor $\sim (1+z)^{2.6}$, making it more difficult to
determine the intrinsic column density, especially for distant GRBs or for GRBs
occurring at large Galactic column densities.  

Finally, we compared the distribution of GRB column densities with known
redshift with theoretical predictions available in the literature finding good
agreement with the expectation (Reichart \& Price 2002) for bursts occurring
in molecular clouds. 

\begin{acknowledgements}
This work is supported at OAB by funding
from ASI on grant number I/R/039/04, at Penn State by NASA contract NAS5-00136
and at the University of Leicester by the PPARC on grant numbers PPA/G/S/00524
and PPA/Z/S/2003/00507. We gratefully acknowledge the contributions of dozens
of members of the Swift team, who helped make this Observatory possible. 
\end{acknowledgements}

\end{document}